\documentclass[prl,reprint,graphicx]{revtex4-1}

\draft 

\usepackage{color}

\usepackage{natbib}
\usepackage[colorlinks=true,linkcolor=blue,citecolor=blue,anchorcolor=blue,urlcolor=blue]{hyperref}

\usepackage{graphicx}
\usepackage{dcolumn}
\usepackage{bm}

\usepackage[utf8]{inputenc}
\usepackage[T1]{fontenc}
\usepackage{mathptmx}

\begin{document}


\title{Towards Compact and Real-Time Terahertz Dual-Comb Spectroscopy Employing a Self-Detection Scheme} 



\author{Hua Li}
\email{hua.li@mail.sim.ac.cn}
\affiliation{Key Laboratory of Terahertz Solid State Technology, Shanghai Institute of Microsystem and Information Technology, Chinese Academy of Sciences, 865 Changning Road, Shanghai 200050, China}
\affiliation{Center of Materials Science and Optoelectronics Engineering, University of Chinese Academy of Sciences, Beijing 100049, China}

\author{Ziping Li}
\affiliation{Key Laboratory of Terahertz Solid State Technology, Shanghai Institute of Microsystem and Information Technology, Chinese Academy of Sciences, 865 Changning Road, Shanghai 200050, China}
\affiliation{Center of Materials Science and Optoelectronics Engineering, University of Chinese Academy of Sciences, Beijing 100049, China}

\author{Wenjian Wan}

\affiliation{Key Laboratory of Terahertz Solid State Technology, Shanghai Institute of Microsystem and Information Technology, Chinese Academy of Sciences, 865 Changning Road, Shanghai 200050, China}

\author{Kang Zhou}
\affiliation{Key Laboratory of Terahertz Solid State Technology, Shanghai Institute of Microsystem and Information Technology, Chinese Academy of Sciences, 865 Changning Road, Shanghai 200050, China}
\affiliation{Center of Materials Science and Optoelectronics Engineering, University of Chinese Academy of Sciences, Beijing 100049, China}

\author{Xiaoyu Liao}
\affiliation{Key Laboratory of Terahertz Solid State Technology, Shanghai Institute of Microsystem and Information Technology, Chinese Academy of Sciences, 865 Changning Road, Shanghai 200050, China}
\affiliation{Center of Materials Science and Optoelectronics Engineering, University of Chinese Academy of Sciences, Beijing 100049, China}

\author{Sijia Yang}
\affiliation{Key Laboratory of Terahertz Solid State Technology, Shanghai Institute of Microsystem and Information Technology, Chinese Academy of Sciences, 865 Changning Road, Shanghai 200050, China}
\affiliation{Center of Materials Science and Optoelectronics Engineering, University of Chinese Academy of Sciences, Beijing 100049, China}

\author{Chenjie Wang}
\affiliation{Key Laboratory of Terahertz Solid State Technology, Shanghai Institute of Microsystem and Information Technology, Chinese Academy of Sciences, 865 Changning Road, Shanghai 200050, China}

\author{J. C. Cao}
\affiliation{Key Laboratory of Terahertz Solid State Technology, Shanghai Institute of Microsystem and Information Technology, Chinese Academy of Sciences, 865 Changning Road, Shanghai 200050, China}
\affiliation{Center of Materials Science and Optoelectronics Engineering, University of Chinese Academy of Sciences, Beijing 100049, China}

\author{Heping Zeng}
\affiliation{State Key Laboratory of Precision Spectroscopy, East China Normal University, Shanghai 200062, China.}

\date{\today}

\begin{abstract}
Due to its fast and high resolution characteristics, dual-comb spectroscopy has attracted an increasing amount of interest since its first demonstration. In the terahertz frequency range where abundant absorption lines (finger prints) of molecules are located, multiheterodyne spectroscopy that employs the dual-comb technique shows an advantage in real-time spectral detection over the traditional Fourier transform infrared or time domain spectroscopies. Here, we demonstrate compact terahertz dual-comb spectroscopy based on quantum cascade lasers (QCLs). In our experiment, two free-running QCLs generate approximately 120 GHz wide combs centered at 4.2 THz, with slightly different repetition frequencies. We observe that $\sim$490 nW terahertz power coupling of one laser into the other suffices for laser-self-detecting the dual-comb spectrum that is registered by a microwave spectrum analyzer. Furthermore, we demonstrate practical terahertz transmission dual-comb spectroscopy with our device, by implementing a short air path at room temperature. Spectra are shown of semiconductor samples and of moist air, the latter allowing rapid monitoring of the relative humidity. Our devices should be readily extendable to perform imaging, microscopy and near-field microscopy in the terahertz regime.
\end{abstract}

\maketitle 


Compared to traditional Fourier transform infrared and time domain spectroscopies \cite{FTIR,TDS1,TDS2,FTIR-TDS}, dual-comb spectroscopy \cite{KeilmannOL2004,KeilmannOE2005,Bernhardt10,DC16} shows advantages not only in fast data acquisition but also in terms of high spectral resolution. In the terahertz frequency range, due to the high power and broad frequency coverage, the electrically pumped terahertz quantum cascade laser (QCL) \cite{1stTHzQCL} is an ideal candidate for dual-comb multiheterodyne spectroscopy.

Although multiheterodyne detection using QCL combs has already been demonstrated using either an on-chip configuration \cite{DCMarkus,Octave} or a laser + fast external detector system \cite{DCFaist,Yang16,WysockiDC}, a compact dual-comb system is still much in demand for practical applications. Concerning the on-chip dual comb, it is indeed compact, but it is incapable of substance detection because the optical coupling between two laser combs is through the edge of the bottom metal for double-metal waveguide QCLs \cite{DCMarkus,Octave} or the laser substrate for single plasmon waveguide QCLs \cite{Lionchip}. Multiheterodyne detection using fast terahertz detectors, such as Schottky mixers \cite{Schottky,Yang16}, superconducting hot-electron bolometers \cite{HEB,Yang16}, or quantum well infrared photodetectors \cite{LiSR}, is exploitable for the detection of molecular finger prints. However, the external fast detector will increase the complexity of the system. The recent trend for the development of terahertz dual-comb spectroscopy is toward compact setups and miniaturization, for example, by using single-cavity mode-locked fiber lasers \cite{HuDCS2018} or electro-optic modulation \cite{JerezDCS2019}. 

In this work, we propose a compact terahertz multiheterodyne spectrometer based on two QCL frequency combs. There are no moving parts in the compact system and only one cryostat is needed. No optical mirror is used for the optical coupling and alignment. Employing the self-detection of QCL combs, multiple dual-comb spectra at different carrier frequencies resulting from free-space optical coupling are successfully obtained. Furthermore, the water absorption and transmission of a GaAs etalon are measured using the multiheterodyne dual-comb system. The microwave-terahertz frequency link is also determined.

\begin{figure}[!t]
	\centering
	\includegraphics[width=0.85\linewidth]{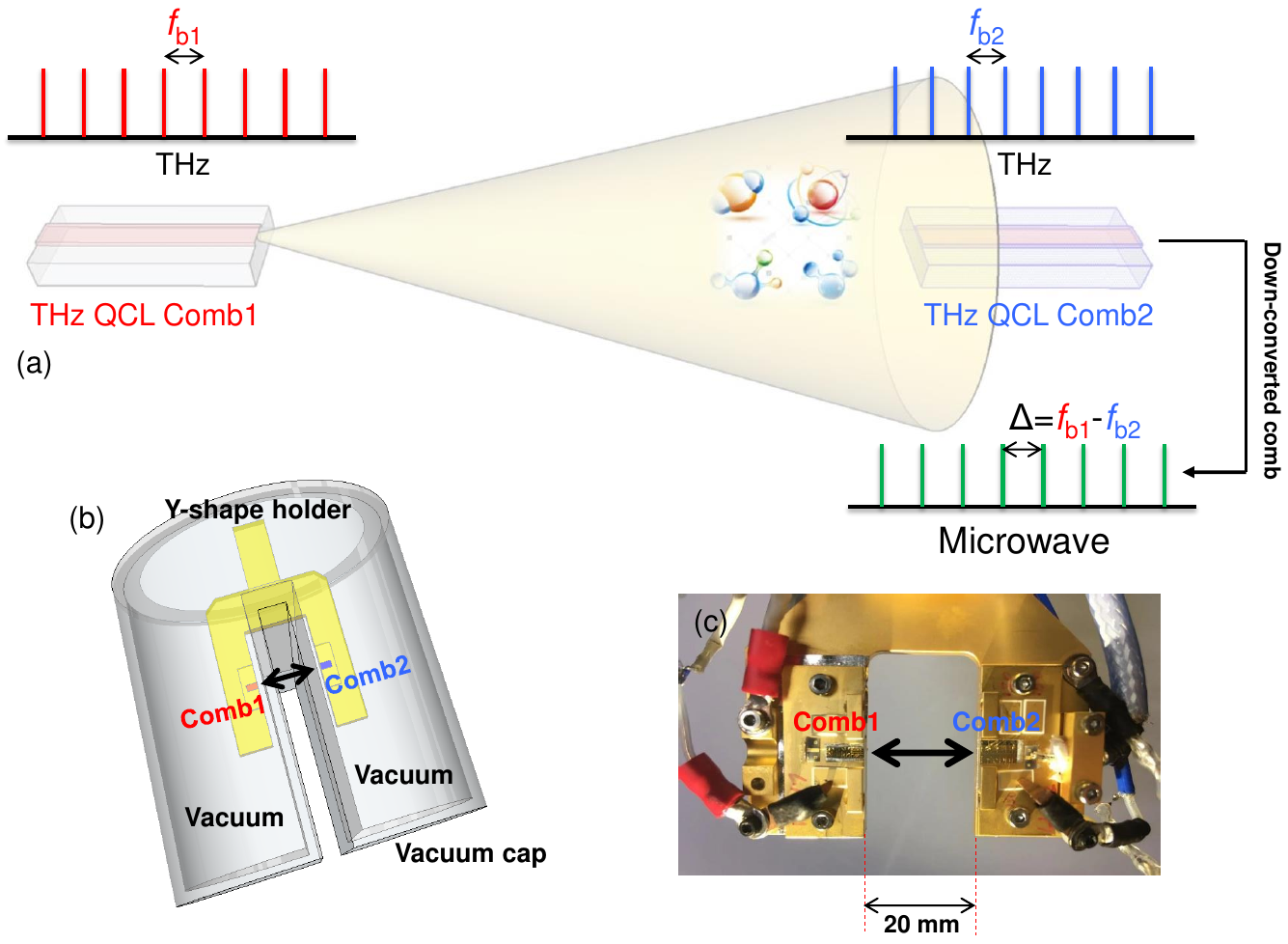}
	\caption{Schematics of the compact terahertz dual-comb multiheterodyne system. (a) Optical coupling of the two terahertz QCL combs and illustration of microwave frequency synthesis. $f$$_{\textrm{b1}}$ and $f$$_{\textrm{b2}}$ denote the intermode beat note frequencies of Comb1 and Comb2, respectively. The molecules in (a) represent samples for the dual-comb spectroscopy. (b) Y-shape dual-comb geometry with a vacuum cap. (c) Photo of the Y-shape holder. The distance between the two output laser facets is 20 mm. The two laser combs have identical dimensions (a 6-mm-long cavity and a 150-$\mu$m-wide ridge). The thick black arrows in (b) and (c) schematically show the terahertz beam propagating directions.}
	\label{Schematic}
\end{figure}

Figure \ref{Schematic}a shows the schematics of the compact terahertz multiheterodyne dual-comb system. The two terahertz QCLs (Comb1 and Comb2) with a hybrid active region (bound-to-continuum for photo emission and resonant-phonon for fast carrier depopulation) \cite{Hybrid} and a single plasmon waveguide are placed face to face. An optimal ridge width of 150 $\mu$m is employed to obtain a flat group velocity dispersion for ideal frequency comb operation \cite{ZhouAPL}. One of the lasers, Comb2, is used as a fast detector for recording the down-converted dual-comb spectra. The QCL can be used as a fast detector due to the fast intersubband relaxation process in the active region \cite{LiOE}. Because the two laser combs are spatially separated, the multiheterodyne system can be used for fast spectral detection by introducing samples between the laser combs. The compact multiheterodyne detection is made possible by judiciously designing a Y-shape holder, as shown in Figure \ref{Schematic}b. The two laser combs sit on the two branches of the holder, with a high density polyethylene vacuum cap designed to create a sample chamber (open air, as shown in Figure \ref{Schematic}b) between the combs. Figure \ref{Schematic}c shows an optical photo of the Y-shape holder with the two laser combs mounted on the branches. The distance between the two output laser facets is 20 mm.

\begin{figure}[!t]
	\centering
	\includegraphics[width=0.95\linewidth]{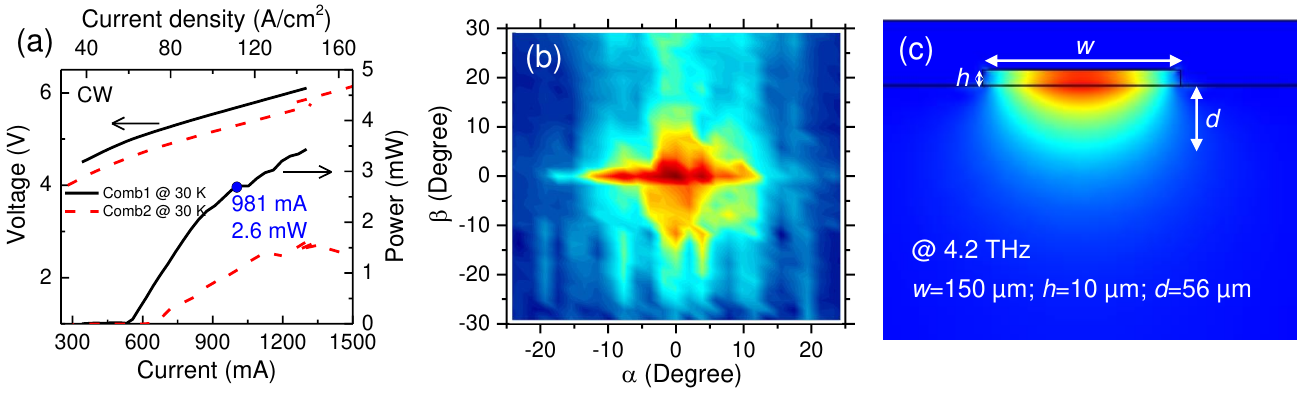}
	\caption{(a) Light-current-voltage characteristics in continuous wave (cw) mode of the two lasers recorded at a stabilized temperature of 30 K. (b) Far-field beam profile of Comb1 measured at 1000 mA. (c) Calculated two-dimensional electric field distribution of the fundamental mode at 4.2 THz for the QCL structure. $w$, $h$, and $d$ denote the ridge width, the ridge height, and the depth into the substrate that the optical mode can reach, respectively.}
	\label{coupling}
\end{figure}

The multiheterodyne system shown in Figure \ref{Schematic} is compact and does not need additional fast detectors and optics for alignments. From the schematic illustration shown in Figure \ref{Schematic}a, we understand that, due to the divergent emission of terahertz QCLs, only a very small portion of the terahertz light emitted from Comb1 finally couples into the detection comb (Comb2). The power that is coupled into Comb2, $P$, can be written as 
\begin{equation}\label{formula1}
P=P_0\times\frac{\int_{\alpha_0}\int_{\beta_0}I(\alpha,\beta)\textrm{d}{\alpha}\textrm{d}{\beta}}{\int_{\textrm{total}}I(\alpha,\beta)\textrm{d}{\alpha}\textrm{d}{\beta}},
\end{equation}
where $P_0$ is the total power of Comb1 that reaches the facet plane of Comb2, $I(\alpha,\beta)$ is the far-field intensity distribution obtained from Figure \ref{coupling}b, $\alpha$ and $\beta$ denote, respectively, horizontal and vertical angles. The numerator in Equation \ref{formula1} is the intensity integral that measures the light interacts with the detector (Comb2). $\alpha_0$ and $\beta_0$ are horizontal and vertical angular ranges obtained by considering the field distribution of the laser comb on the facet (see Figure \ref{coupling}c) and a 20-mm distance between the two laser combs. The denominator in Equation \ref{formula1} depicts the total far-field intensity. By considering a 50\% loss (window transmission and water absorption), $P_0$ can be estimated to be 1.3 mW (see Figure \ref{coupling}a for the measured optical power of Comb1). From Figure \ref{coupling}a, we can find that even the two lasers have similar dimensions, the electrical and optical characteristics are still not the same. This is because during the fabrication and mounting processes, imperfections such as the nonuniform etching of the ridge and defects on the cleaved laser facets can affect the laser performance. From the horizontal and vertical field dimensions shown in Figure \ref{coupling}c, $\alpha_0$ is calculated to be between -0.09$^\circ$ and +0.09$^\circ$, and similarly $\beta_0$ is between -0.21$^\circ$ and +0.21$^\circ$. The ratio $R$=$\frac{\int_{\alpha_0}\int_{\beta_0}I(\alpha,\beta)\rm{d}{\alpha}\rm{d}{\beta}}{\int_{\rm{total}}I(\alpha,\beta)\rm{d}{\alpha}\rm{d}{\beta}}$ is then calculated to be 3.81$\times$10$^{-4}$. Finally, we are able to estimate that the power $P$ that is injected into Comb2 for generating the multiheterodyne beats is $\sim$490 nW. Note that the coupled power $P$ is estimated by considering the practical far-field beam pattern that was measured with an angle step of 1$^\circ$. To calculate the integral in Equation \ref{formula1}, an interpolation method was used to mesh the original experimental data into a two-dimensional data with an angle step of 0.01$^\circ$. Actually, the coupled power is related to the optical injection locking in lasers. Our tests for the on-chip dual-comb devices show that as the two single plasmon QCLs are spatially separated by less than 400 $\mu$m, optical injection locking occurs, which results in no difference between $f$$_{\textrm{b1}}$ and $f$$_{\textrm{b2}}$. Once the two lasers are optically locked with each other, no dual-comb spectra can be observed. Similar to the on-chip dual-comb configuration \cite{Lionchip}, for the current setup employing the same self-detection scheme, strong optical coupling between the combs should be avoided due to the optical injection-locking effect, which is also the reason why optics are not used in this work to improve the optical coupling.

\begin{figure}[!b]
	\centering
	\includegraphics[width=0.85\linewidth]{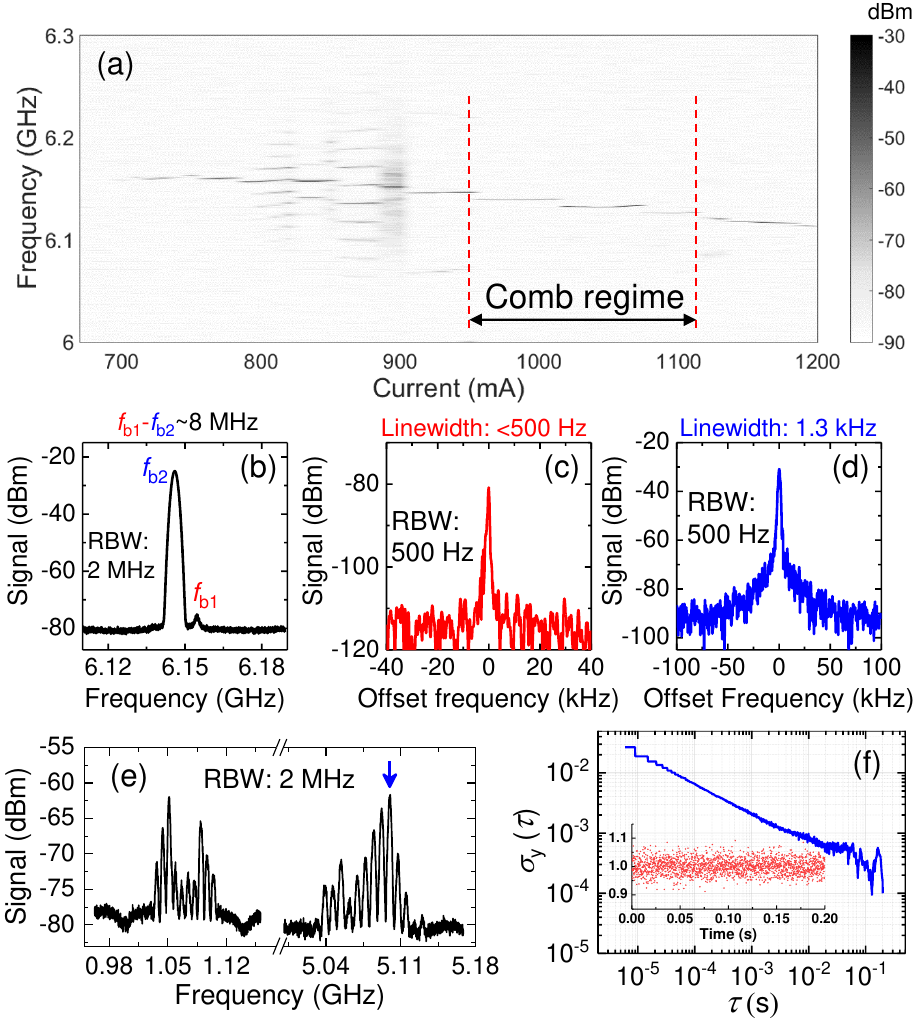}
	\caption{(a) Intermode beat note mapping of laser Comb2 measured with an RBW of 100 kHz. (b) Intermode beat notes measured from the detection comb (Comb2) with an RBW of 2 MHz when the two laser combs are simultaneously switched on. (c), (d) The measured linewidths of $f$$_{\textrm{b1}}$ and $f$$_{\textrm{b2}}$, respectively. (e) Representative dual-comb spectra at approximately 1.07 and 5.08 GHz. (f) Allan deviation of the amplitudes of the dual-comb line, as indicated by the blue arrow in (e). The inset shows the oscillations of the amplitude of the dual-comb line. The data shown in (b)-(f) were recorded as QCL Comb1 and Comb2 operated at 1000 and 945 mA, respectively, in cw at a heat sink temperature of 20 K.}
	\label{dual-comb}
\end{figure}

Before the dual-comb operation, we first evaluate the comb characteristics of one single laser comb by electrically measuring the intermode beat note using the laser itself as a detector \cite{Gellie10,LiOE,DCMarkus}. 
Figure \ref{dual-comb}a plots the intermode beat note mapping of Comb2 to show the full picture of the comb behavior in the entire current dynamic range. As the current is below 900 mA, the intermode beat note spectra are characterized by either broad signal or multiline structures, which indicates that the laser is not working as a comb in this current regime. Frequency comb operation characterized by a single narrow intermode beat note is obtained in the current range between 950 and 1100 mA, which is marked by the two vertical dashed lines. The intermode beat note mapping of Comb1 is also measured and shown in Figure S2 (Supporting Information). Note that Comb1 is not mounted with a microwave transmission line. To measure its intermode beat note signal, we simply place the RF probe close to the laser chip. Because there is no physical connection between the laser and the RF probe, the measured signal is much weaker (45 dB) than the signal measured from Comb2. However, the comb operation of Comb1 is also perfectly obtained. To ensure frequency comb operation, the two laser combs are driven in the current range between 950 and 1100 mA.

For multiheterodyne dual-comb detection, similarly to the intermode beat note measurement shown in Figure \ref{dual-comb}a, we employ the QCL self-detection scheme (details of the dual-comb measurement setup are shown in the Supporting Information, Figure S1). The two laser combs are electrically driven by two different DC power supplies. The two laser combs emit terahertz light simultaneously and the light from one comb will couple into the other comb via the free-space. The multiheterodyne beating signal is detected by Comb2 and the dual-comb spectrum can be recorded in real-time using a spectrum analyzer with the assistance of a microwave amplifier. 

To successfully obtain the downconverted dual-comb spectra, the two laser combs should demonstrate narrow beat note lines with a slight frequency difference ($\Delta{f}$=$| f_{\textrm{b1}}$-$f_{\textrm{b2}}|$) which determines the line spacing of the downconverted dual-comb spectrum. In Figure \ref{dual-comb}b, we show the intermode beat note signal measured when Comb1 and Comb2 are simultaneously operated at 1000 and 945 mA in cw mode, respectively. Because the signal is measured via the microwave transmission line, which is electrically connected to the detection comb (Comb2), the signal at $f_{\textrm{b2}}$ is much stronger than that at $f_{\textrm{b1}}$. The frequency difference between $f_{\textrm{b1}}$ and $f_{\textrm{b2}}$ ($\Delta{f}$) is measured to be 8 MHz which is equal to the measured line spacing of the downconverted dual-comb spectra shown in Figure \ref{dual-comb}e. The high resolution beat note spectra for Comb1 and Comb2 are shown in Figures \ref{dual-comb}c and \ref{dual-comb}d, respectively, as measured with a resolution bandwidth (RBW) of 500 Hz. The measured linewidths for both  signals are comparable; i.e., they are $<$500 Hz and 1.3 kHz for $f_{\textrm{b1}}$ and $f_{\textrm{b2}}$, respectively.

The dual-comb generation results from the multiheterodyne beatings of the two sets of terahertz modes. Different beating processes can generate dual-comb spectra at different carrier frequencies in the microwave range (the detailed beating processes are schematically described in the Supporting Information, Figure S3a). Each downconverted dual-comb line is from the beating between one terahertz mode of Comb1 and one terahertz mode of Comb2. Note that the beatings of the neighboring terahertz modes and the beatings between modes that are far separated can generate dual-comb signals. In this work, four dual-comb spectra up to 11 GHz were successfully observed (see the Supporting Information, Figure S3b). Due to differences in the nonlinear mixing strengths of the terahertz modes and the signal attenuations of the microwave cables at different frequencies, the dual-comb line number decreases as the frequency is increased. Figure \ref{dual-comb}e shows two representative dual-comb spectra at lower carrier frequencies at approximately 1.07 and 5.08 GHz generated from processes 1 and 2, respectively (see the Supporting Information, Figure S3a). We can see that the two dual-comb spectra shown in Figure \ref{dual-comb}e almost mirror each other with respect to the frequency between them [(1.07+5.08)/2=3.075 GHz]. This mirror effect directly shows that the two dual-comb spectra have opposite microwave-frequency links (see the Supporting Information, Figure S4), which also proves that the spectra shown in Figure \ref{dual-comb}e are dual-comb signals resulting from the multiheterodyne beatings. Concerning the generation of downconverted dual-comb spectra, at some drive currents, one group of dual-comb lines can overlap with another group of lines. This aliasing can be intentionally avoided by carefully tuning the currents of the two laser combs. 

To further demonstrate the free-space optical coupling and real-time detection characteristics of the dual-comb system, we recorded a video to show that the dual-comb spectra with an A4 paper or metal board in the beam path can be obtained in real time (see the Supplementary Video).  

Noise properties are critical for a dual-comb system, determining how stable a system is for practical measurements. To evaluate the noise properties of the current dual-comb system, an Allan deviation analysis of the amplitudes is performed for the dual-comb line indicated by the blue arrow in Figure \ref{dual-comb}e. The amplitudes of the selected RF line were measured with a 2 $\mu$s time resolution for a total acquisition time of 200 ms. The Allan deviation of the normalized amplitude oscillations (see the inset of Figure \ref{dual-comb}f) as a function of the integration time is shown in Figure \ref{dual-comb}f. The Allan deviation decreases monotonically with the integration time when the latter is shorter than 20 ms. When the integration time is longer than 20 ms, oscillations in the Allan deviation can be observed. However, these oscillations do not show up for other Allan deviation measurements with a longer acquisition time of 1 s (see the Supporting Information, Figure S6). Therefore, we attribute the oscillating behavior in Figure \ref{dual-comb}f to some mechanical vibration noise during the measurement. We also compare the amplitude noise with those obtained from other dual-comb systems based on QCLs. At an integration time of 1 ms, we obtained an Allan deviation at the 10$^{-3}$ level similar to that measured from a mid-infrared QCL dual-comb system \cite{DCFaist}. At an integration time of 100 ms, our amplitude noise is two orders lower than that measured from a terahertz QCL dual-comb system \cite{WysockiDC}.

\begin{figure}[!t]
	\centering
	\includegraphics[width=0.9\linewidth]{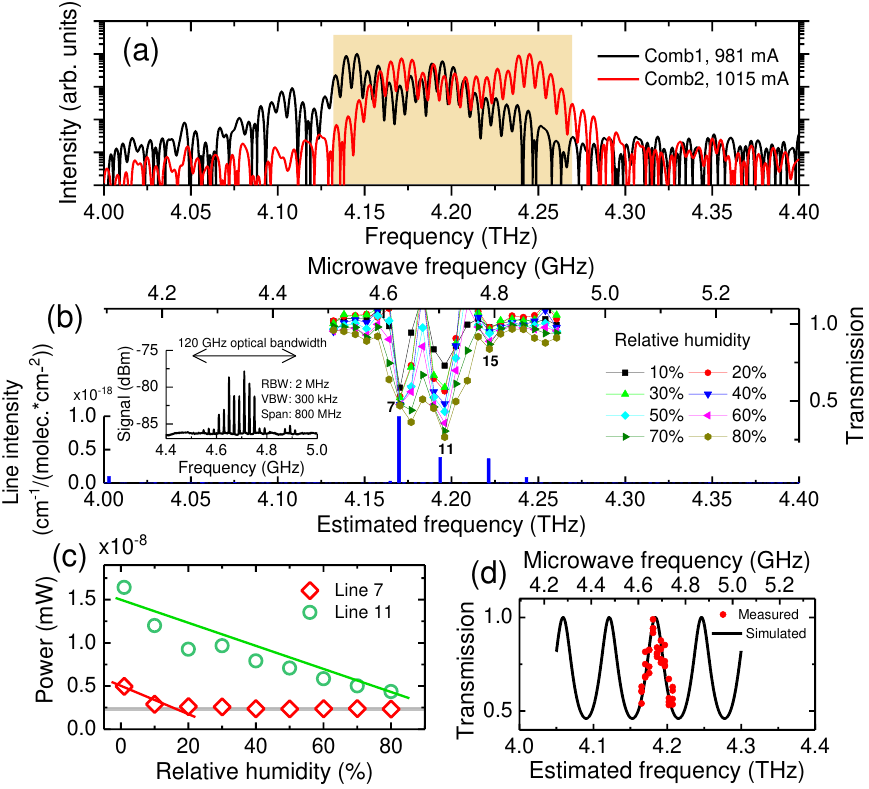}
	\caption{(a) Emission spectra of the QCL Comb1 (black) and Comb2 (red). The shaded area depicts the frequency regime where the two laser combs are spectrally overlapped for dual-comb generations. (b) Peak intensities of dual-comb spectra at different RH values. All the data were normalized to the results measured at 1\% RH. The numbers 7, 11, and 15 denote the line index of the dual-comb spectra that correspond to the three water absorptions at 4.17, 4.193, and 4.221 THz, respectively. The water absorption lines obtained from the HITRAN database are also shown in (b) for an approximate frequency calibration from microwave to terahertz. The inset shows a typical dual-comb spectrum measured with an RH of 1\%. (c) Power as a function of RH for line 7 and 11. The scatters are measured results, and the solid lines show the linear relationship between the measured power and the RH. The horizontal line shows the noise level of the dual-comb measurement system. (d) Transmission of a 625-$\mu$m-thick GaAs etalon measured using the dual-comb system. The circles are the extracted peak intensities from the dual-comb lines. The black curve is a simulation for comparison. All the data were recorded when Comb1 and Comb2 were operated at 981 and 1015 mA, respectively, in cw at a stabilized heat sink temperature of 34.5 K.}
	\label{water}
\end{figure}

In Figure \ref{water}, we verify that the multiheterodyne system is capable of spectroscopic applications. Figure \ref{water}a shows the terahertz emission spectra of Comb1 and Comb2 driven at 981 and 1015 mA, respectively, at a stabilized heat sink temperature of 34.5 K. Note that the pumping conditions and operation temperature of the two laser combs for the measurement shown in Figure \ref{water} are not the same as those used in Figure \ref{dual-comb}. Therefore, the dual-comb bandwidth, line number, and dual-comb central frequency are different from those reported in Figure \ref{dual-comb}. The shaded area in Figure \ref{water}a sketches the frequency range where the two combs strongly interact with each other to generate dual-comb spectra (see the inset of Figure \ref{water}b). The first spectroscopy is performed with a sample of water vapor. We change the relative humidity (RH) of the sample chamber and measure the dual-comb spectra with an optical bandwidth of 120 GHz at a frequency of approximately 4.7 GHz, as shown in the inset of Figure \ref{water}b. The dual-comb spectra are recorded directly using a spectrum analyzer with an RBW of 2 MHz, a video bandwidth (VBW) of 300 kHz, and a span of 800 MHz (see the Supporting Information for details of the dual-comb measurement). Then, the peaks are taken from the dual-comb spectra that were measured at different RH values for all 21 lines. The transmission plotted in Figure \ref{water}b is obtained by normalizing the peaks to those measured at 1\% RH. We can clearly observe two strong water absorptions at 4.63 and 4.71 GHz in the microwave regime marked by lines 7 and 11, respectively. To roughly calibrate the frequency from microwave to terahertz, the water absorption lines extracted from the HITRAN database are also plotted as blue lines in Figure \ref{water}b. The estimated terahertz frequencies for the dual-comb lines from modes 1 to 21 are between 4.139 to 4.261 THz, covering an optical bandwidth of 122 GHz, which agrees well with the frequency range shown in Figure \ref{water}a for generating dual-comb spectra. We find that lines 7 and 11 well correspond to the two strong water absorptions at 4.17 and 4.193 THz, respectively. Because the absorption at 4.17 THz is three times stronger than the one at 4.193 THz, the dual-comb absorption marked by line 7 should, in principle, be deeper, which contradicts the measured result. This is due to the signal-to-noise ratio limit of the dual-comb measurement system. The multiheterodyne dual-comb spectra measured at different RH values are plotted in Figure S7 (Supporting Information). Due to the strong absorption at 4.17 THz, when the RH is larger than 20\%, line 7 becomes very weak, and the power drops into the noise level. Therefore, we see that the peak at line 7 ``saturates" when the RH is larger than 20\%. Figure \ref{water}c shows the RH dependence of the power for lines 7 (red diamonds) and 11 (green circles). It can be clearly seen that the power of line 11 has a linear relationship with the RH, which follows Lambert's law for water absorption \cite{RH1,RH2}. The linear relationship can be employed to measure and calibrate the RH values of water vapor. However, the power of line 7 drops dramatically into the noise level (2.17-2.48$\times$10$^{-9}$ mW, as indicated by the horizontal gray line), which is extracted from the dual-comb spectra shown in Figure S7. Limited by the signal-to-noise ratio, the linear dependence between the line power and the RH cannot be observed in the entire RH range for line 7. Here, it is worth mentioning that other than lines 7 and 11 in the dual-comb transmission plot shown in Figure \ref{water}b, another deep spectrum marked by line 15 can also be observed, corresponding to the water absorption at 4.221 THz. Due to the same signal-to-noise ratio limit (see Figure S7, Supporting Information), the water absorption marked by line 15 is less pronounced. 

We also employ the dual-comb system to measure the transmission of a GaAs etalon to further verify its ability for spectroscopy. Figure \ref{water}d shows the measured transmission of a 625-$\mu$m-thick GaAs etalon (red circles) together with the simulated transmission (black curve). The experimental data shown in Figure \ref{water}d are obtained by taking ratios of the peak intensities of the dual-comb spectra measured with and without the GaAs etalon. The estimated terahertz frequency is obtained by using the two water absorption lines shown in Figure \ref{water}b. The simulation is performed by employing a transfer matrix method \cite{LiCC}. Unfortunately, due to the signal attenuation introduced by the thick sample, the effective optical bandwidth for this measurement is approximately 50 GHz from 4.15 to 4.20 THz. However, the agreement between the experimental data and the simulation verifies that the multiheterodyne system is capable of spectroscopies. To further prove its validity for spectroscopy, we also measure the dual-comb transmission for another thinner GaAs etalon with a thickness of 200 $\mu$m (see Figure S9, Supporting Information). When the thickness is reduced to 200 $\mu$m, the spacing between two transmission maxima increases. Although the dual-comb transmission cannot depict the full picture of the transmission with frequency due to the limited optical bandwidth, the simulation qualitatively fits the measured results.

Optical feedback is a critical issue for the frequency comb and dual-comb operations. In general, the feedback can affect the comb states, especially when an etalon sample is placed in our dual-comb system. Because of this, we investigate the feedback effect in the current dual-comb system. A two-dimensional finite-element simulation is employed to quantify the power reflected back from the etalon to the laser cavity. Details of the simulation can be found in the Supporting Information. Figure S10 (Supporting Information) shows the calculated electric field distribution and the feedback power ratio ($P_{\rm{r}}$/$P_{\rm{i}}$, here $P_{\rm{i}}$ is the injection power at the laser emitting facet and $P_{\rm{r}}$ is the measured reflected power at the port that is close to the laser facet) as a function of the distance between the laser and the etalon. It has been found that without any optics for alignment, the power that is reflected back to the laser cavity is very small ($\sim$10$^{-10}$ feedback power ratio). Because of the weak feedback, the frequency comb operation is not destroyed during the etalon transmission measurement. 

In the current dual-comb system, the other laser, Comb1, can also be used for multiheterodyne detection. If both dual-comb signals detected by Comb1 and Comb2 can be acquired simultaneously, the noise can be significantly reduced. However, as we already explained, because there is no physical wire connection between the laser chip and the RF probe for the mounting of Comb1, the measured RF signal is much weaker than that measured from Comb2. From the data shown in Figures \ref{dual-comb}d and \ref{dual-comb}e where Comb2 is used as a detector, the dual-comb signal is $\sim$30 dB weaker than the intermode beat note signal. We can simply assume that when Comb1 is used as a detector, the measured power difference between the dual-comb and intermode beat note signals is 30 dB. From Figure S2, we can roughly calculate the power of the dual-comb signal measured from Comb1 (i.e., -105 dBm), which is far below the noise level of the dual-comb measurement system (-86 dBm, as shown in the inset of Figure \ref{water}b and measured with an RBW of 2 MHz and a VBW of 300 kHz). Furthermore, as shown in Figure \ref{coupling}a, the output power of Comb2 is much smaller than that of Comb1, and therefore, the coupled power into Comb1 for multiheterodyne detection is also lower. Both factors mentioned above make it difficult to measure the dual-comb signals simultaneously to reduce the noise. In the current dual-comb system, an effective way to reduce the noise is to increase the integration time for data acquisition.

In Figures \ref{water}(b) and \ref{water}(d), the estimated optical frequency is given in THz. Note that the microwave-terahertz frequency link depends on the relative line positions of the two sets of terahertz comb lines and the relative magnitudes of $f$$_\textrm{b1}$ and $f$$_\textrm{b2}$ (see the Supporting Information, Figure S4). The mirrored feature of the dual-comb spectra at approximately 1.07 and 5.08 GHz shown in Figure \ref{dual-comb}e indicates that the two sets of dual-comb lines have an opposite microwave-terahertz frequency link. However, because the relative line positions of the two sets of terahertz modes are unknown in the current free-running dual-comb system, it is difficult to determine the specific frequency link for the dual-comb spectra; i.e., we cannot establish wether the increase in microwave frequency corresponds to either an increase or a decrease in the terahertz frequency. In our dual-comb system, the water absorption lines given by the HITRAN data and the transmission measurement of the GaAs etalon shown in Figure \ref{water} can be used to determine such microwave-terahertz frequency link. We finally find that for the dual-comb spectra at approximately 5.08 GHz (1.07 GHz), the increase in the microwave frequency corresponds to an increase (decrease) in the terahertz frequency. To firmly prove that the microwave-terahertz frequency link shown in Figure \ref{water} is correct, the transmission spectra of moist air and GaAs etalon are replotted by flipping the terahertz frequency as shown in Figure S5 (see the Supporting Information). We find that when we change the frequency link by flipping the terahertz frequency, the measured transmission of the 625-um-thick GaAs etalon cannot match with the calculated result. Therefore, the correctness of the frequency link shown in Figure \ref{water} is further proved. Once the microwave-terahertz frequency link is calibrated, it can be directly used for other measurements.

Although the dual-comb system proposed in this work demonstrates the capability of real-time terahertz dual-comb spectroscopy using the self-detection scheme, it is currently not suitable for high precision applications because external frequency stabilization is not applied. The ambient water absorptions are normally $\sim$1 GHz wide and the long-term (3 minutes) frequency drift of our laser combs is at the hundred kHz level, which is measured by employing the ``max-hold" function of a spectrum analyzer \cite{LiAS}. Therefore, the frequency stability of the dual-comb system is sufficient for the spectroscopic measurement of ambient water absorptions. It is worth noting that in the spectroscopic measurement shown in Figure \ref{water}, an RBW of 2 MHz is used to record the dual-comb traces, and we do not pay much attention to the spectral resolution. The second group of multiheterodyne modes is chosen for spectroscopy simply because this group shows more lines with a higher and more evenly distributed power (see Figure \ref{water}b and Figure S8, Supporting Information). It is true that the second group of multiheterodyne lines results from the beating of terahertz modes that are more separated compared to the first group of lines, which can play an important role in high resolution spectroscopy. To demonstrate that the other groups of dual-comb lines can also be used for spectroscopy, the transmission of water vapors at different RH values measured using the first and third group of lines are shown in Figures S8a and S8b (Supporting Information), respectively. However, due to the smaller number of lines and the poor signal-to-noise ratio, the results shown in Figure S8 are not as good as those measured using the second group of lines. In the future, there is still some work to do to optimize the dual-comb system for high resolution applications. First, the frequency stability and optical bandwidth can, in principle, be significantly improved by external radio frequency (RF) injection and phase locking techniques \cite{Gellie10,BarbieriNP,LiOE}. Furthermore, the RF injection can also tune the terahertz frequency for high resolution spectroscopy in combination with the drive current/voltage tuning techniques. For instance, by sweeping one of the combs with the tuning techniques and fixing the other comb, a higher spectral resolution is definitely feasible with this compact system. Second, to improve the signal-to-noise ratio of the multiheterodyne lines, the distance between the two laser combs, which determines the coupling strength of the two combs, can be optimized to find a balance between optical coupling and injection locking. In addition, some small optics (such as an aspheric silicon lens or off-axis parabolic mirrors) can be used to improve the optical coupling between the two lasers, which can further improve the signal-to-noise ratio of the present dual-comb system. Third, the relative phase relationship between the two sets of modes could be further locked to increase the stability of the system. This can be achieved by sending one of the dual-comb lines to a phase lock loop module and then locking all the multiheterodyne lines without locking the carriers of the combs. Finally, for a controllable terahertz spectroscopy, a gas cell with pressure control is ideal for more accurate measurements. However, the addition of a gas cell is problematic in the current dual-comb configuration due to the limited working distance between the two laser combs.

In summary, we have demonstrated a compact and real-time terahertz dual-comb system that employs a self-detection scheme based on QCLs. The real-time multiheterodyne dual-comb was successfully obtained with $\sim$490 nW terahertz light coupled into the detection comb. The multiheterodyne system with an optical bandwidth of 120 GHz shows the ability to perform ambient water absorption and etalon transmission measurements. The portable system can be potentially implemented in various application scenarios for fast spectroscopy and/or toxic substance identifications.

The authors thank Dr. Fritz Keilmann for fruitful discussions on the dual-comb technology. This work is supported by the Fundamental Frontier Scientific Research Program of the Chinese Academy of Sciences (ZDBS-LY-JSC009), the National Natural Science Foundation of China (61875220, 61575214, 61927813, 61404150, 61405233, 61704181, and 11727812), and the Major National Development Project of Scientific Instrument and Equipment (2017YFF0106302).

\bibliography{ref}

\begin{thebibliography}{29}%
\makeatletter
\providecommand \@ifxundefined [1]{%
 \@ifx{#1\undefined}
}%
\providecommand \@ifnum [1]{%
 \ifnum #1\expandafter \@firstoftwo
 \else \expandafter \@secondoftwo
 \fi
}%
\providecommand \@ifx [1]{%
 \ifx #1\expandafter \@firstoftwo
 \else \expandafter \@secondoftwo
 \fi
}%
\providecommand \natexlab [1]{#1}%
\providecommand \enquote  [1]{``#1''}%
\providecommand \bibnamefont  [1]{#1}%
\providecommand \bibfnamefont [1]{#1}%
\providecommand \citenamefont [1]{#1}%
\providecommand \href@noop [0]{\@secondoftwo}%
\providecommand \href [0]{\begingroup \@sanitize@url \@href}%
\providecommand \@href[1]{\@@startlink{#1}\@@href}%
\providecommand \@@href[1]{\endgroup#1\@@endlink}%
\providecommand \@sanitize@url [0]{\catcode `\\12\catcode `\$12\catcode
  `\&12\catcode `\#12\catcode `\^12\catcode `\_12\catcode `\%12\relax}%
\providecommand \@@startlink[1]{}%
\providecommand \@@endlink[0]{}%
\providecommand \url  [0]{\begingroup\@sanitize@url \@url }%
\providecommand \@url [1]{\endgroup\@href {#1}{\urlprefix }}%
\providecommand \urlprefix  [0]{URL }%
\providecommand \Eprint [0]{\href }%
\providecommand \doibase [0]{http://dx.doi.org/}%
\providecommand \selectlanguage [0]{\@gobble}%
\providecommand \bibinfo  [0]{\@secondoftwo}%
\providecommand \bibfield  [0]{\@secondoftwo}%
\providecommand \translation [1]{[#1]}%
\providecommand \BibitemOpen [0]{}%
\providecommand \bibitemStop [0]{}%
\providecommand \bibitemNoStop [0]{.\EOS\space}%
\providecommand \EOS [0]{\spacefactor3000\relax}%
\providecommand \BibitemShut  [1]{\csname bibitem#1\endcsname}%
\let\auto@bib@innerbib\@empty
\bibitem [{\citenamefont {Coleman}\ and\ \citenamefont {Painter}(1984)}]{FTIR}%
  \BibitemOpen
  \bibfield  {author} {\bibinfo {author} {\bibfnamefont {M.~M.}\ \bibnamefont
  {Coleman}}\ and\ \bibinfo {author} {\bibfnamefont {P.~C.}\ \bibnamefont
  {Painter}},\ }\href {<Go to ISI>://WOS:A1984AEU5700001} {\bibfield  {journal}
  {\bibinfo  {journal} {Applied Spectroscopy Reviews}\ }\textbf {\bibinfo
  {volume} {20}},\ \bibinfo {pages} {255} (\bibinfo {year} {1984})}\BibitemShut
  {NoStop}%
\bibitem [{\citenamefont {Vanexter}\ \emph {et~al.}(1989)\citenamefont
  {Vanexter}, \citenamefont {Fattinger},\ and\ \citenamefont
  {Grischkowsky}}]{TDS1}%
  \BibitemOpen
  \bibfield  {author} {\bibinfo {author} {\bibfnamefont {M.}~\bibnamefont
  {Vanexter}}, \bibinfo {author} {\bibfnamefont {C.}~\bibnamefont {Fattinger}},
  \ and\ \bibinfo {author} {\bibfnamefont {D.}~\bibnamefont {Grischkowsky}},\
  }\href {<Go to ISI>://WOS:A1989AV56600012} {\bibfield  {journal} {\bibinfo
  {journal} {Optics Letters}\ }\textbf {\bibinfo {volume} {14}},\ \bibinfo
  {pages} {1128} (\bibinfo {year} {1989})}\BibitemShut {NoStop}%
\bibitem [{\citenamefont {Ho}\ \emph {et~al.}(2010)\citenamefont {Ho},
  \citenamefont {Guo},\ and\ \citenamefont {Zhang}}]{TDS2}%
  \BibitemOpen
  \bibfield  {author} {\bibinfo {author} {\bibfnamefont {I.~C.}\ \bibnamefont
  {Ho}}, \bibinfo {author} {\bibfnamefont {X.~Y.}\ \bibnamefont {Guo}}, \ and\
  \bibinfo {author} {\bibfnamefont {X.~C.}\ \bibnamefont {Zhang}},\ }\href {<Go
  to ISI>://WOS:000274791200109} {\bibfield  {journal} {\bibinfo  {journal}
  {Optics Express}\ }\textbf {\bibinfo {volume} {18}},\ \bibinfo {pages} {2872}
  (\bibinfo {year} {2010})}\BibitemShut {NoStop}%
\bibitem [{\citenamefont {Han}\ \emph {et~al.}(2001)\citenamefont {Han},
  \citenamefont {Tani}, \citenamefont {Usami}, \citenamefont {Kono},
  \citenamefont {Kersting},\ and\ \citenamefont {Zhang}}]{FTIR-TDS}%
  \BibitemOpen
  \bibfield  {author} {\bibinfo {author} {\bibfnamefont {P.~Y.}\ \bibnamefont
  {Han}}, \bibinfo {author} {\bibfnamefont {M.}~\bibnamefont {Tani}}, \bibinfo
  {author} {\bibfnamefont {M.}~\bibnamefont {Usami}}, \bibinfo {author}
  {\bibfnamefont {S.}~\bibnamefont {Kono}}, \bibinfo {author} {\bibfnamefont
  {R.}~\bibnamefont {Kersting}}, \ and\ \bibinfo {author} {\bibfnamefont
  {X.~C.}\ \bibnamefont {Zhang}},\ }\href {<Go to ISI>://WOS:000166688300057}
  {\bibfield  {journal} {\bibinfo  {journal} {Journal of Applied Physics}\
  }\textbf {\bibinfo {volume} {89}},\ \bibinfo {pages} {2357} (\bibinfo {year}
  {2001})}\BibitemShut {NoStop}%
\bibitem [{\citenamefont {Keilmann}\ \emph {et~al.}(2004)\citenamefont
  {Keilmann}, \citenamefont {Gohle},\ and\ \citenamefont
  {Holzwarth}}]{KeilmannOL2004}%
  \BibitemOpen
  \bibfield  {author} {\bibinfo {author} {\bibfnamefont {F.}~\bibnamefont
  {Keilmann}}, \bibinfo {author} {\bibfnamefont {C.}~\bibnamefont {Gohle}}, \
  and\ \bibinfo {author} {\bibfnamefont {R.}~\bibnamefont {Holzwarth}},\ }\href
  {\doibase Doi 10.1364/Ol.29.001542} {\bibfield  {journal} {\bibinfo
  {journal} {Optics Letters}\ }\textbf {\bibinfo {volume} {29}},\ \bibinfo
  {pages} {1542} (\bibinfo {year} {2004})}\BibitemShut {NoStop}%
\bibitem [{\citenamefont {Schliesser}\ \emph {et~al.}(2005)\citenamefont
  {Schliesser}, \citenamefont {Brehm}, \citenamefont {Keilmann},\ and\
  \citenamefont {van~der Weide}}]{KeilmannOE2005}%
  \BibitemOpen
  \bibfield  {author} {\bibinfo {author} {\bibfnamefont {A.}~\bibnamefont
  {Schliesser}}, \bibinfo {author} {\bibfnamefont {M.}~\bibnamefont {Brehm}},
  \bibinfo {author} {\bibfnamefont {F.}~\bibnamefont {Keilmann}}, \ and\
  \bibinfo {author} {\bibfnamefont {D.~W.}\ \bibnamefont {van~der Weide}},\
  }\href {\doibase Doi 10.1364/Opex.13.009029} {\bibfield  {journal} {\bibinfo
  {journal} {Optics Express}\ }\textbf {\bibinfo {volume} {13}},\ \bibinfo
  {pages} {9029} (\bibinfo {year} {2005})}\BibitemShut {NoStop}%
\bibitem [{\citenamefont {Bernhardt}\ \emph {et~al.}(2010)\citenamefont
  {Bernhardt}, \citenamefont {Ozawa}, \citenamefont {Jacquet}, \citenamefont
  {Jacquey}, \citenamefont {Kobayashi}, \citenamefont {Udem}, \citenamefont
  {Holzwarth}, \citenamefont {Guelachvili}, \citenamefont {H\"{a}nsch},\ and\
  \citenamefont {Picque}}]{Bernhardt10}%
  \BibitemOpen
  \bibfield  {author} {\bibinfo {author} {\bibfnamefont {B.}~\bibnamefont
  {Bernhardt}}, \bibinfo {author} {\bibfnamefont {A.}~\bibnamefont {Ozawa}},
  \bibinfo {author} {\bibfnamefont {P.}~\bibnamefont {Jacquet}}, \bibinfo
  {author} {\bibfnamefont {M.}~\bibnamefont {Jacquey}}, \bibinfo {author}
  {\bibfnamefont {Y.}~\bibnamefont {Kobayashi}}, \bibinfo {author}
  {\bibfnamefont {T.}~\bibnamefont {Udem}}, \bibinfo {author} {\bibfnamefont
  {R.}~\bibnamefont {Holzwarth}}, \bibinfo {author} {\bibfnamefont
  {G.}~\bibnamefont {Guelachvili}}, \bibinfo {author} {\bibfnamefont {T.~W.}\
  \bibnamefont {H\"{a}nsch}}, \ and\ \bibinfo {author} {\bibfnamefont
  {N.}~\bibnamefont {Picque}},\ }\href {<Go to ISI>://WOS:000273710700025}
  {\bibfield  {journal} {\bibinfo  {journal} {Nature Photonics}\ }\textbf
  {\bibinfo {volume} {4}},\ \bibinfo {pages} {55} (\bibinfo {year}
  {2010})}\BibitemShut {NoStop}%
\bibitem [{\citenamefont {Coddington}\ \emph {et~al.}(2016)\citenamefont
  {Coddington}, \citenamefont {Newbury},\ and\ \citenamefont {Swann}}]{DC16}%
  \BibitemOpen
  \bibfield  {author} {\bibinfo {author} {\bibfnamefont {I.}~\bibnamefont
  {Coddington}}, \bibinfo {author} {\bibfnamefont {N.}~\bibnamefont {Newbury}},
  \ and\ \bibinfo {author} {\bibfnamefont {W.}~\bibnamefont {Swann}},\
  }\href@noop {} {\bibfield  {journal} {\bibinfo  {journal} {Optica}\ }\textbf
  {\bibinfo {volume} {3}},\ \bibinfo {pages} {414} (\bibinfo {year}
  {2016})}\BibitemShut {NoStop}%
\bibitem [{\citenamefont {K\"{o}hler}\ \emph {et~al.}(2002)\citenamefont
  {K\"{o}hler}, \citenamefont {Tredicucci}, \citenamefont {Beltram},
  \citenamefont {Beere}, \citenamefont {Linfield}, \citenamefont {Davies},
  \citenamefont {Ritchie}, \citenamefont {Iotti},\ and\ \citenamefont
  {Rossi}}]{1stTHzQCL}%
  \BibitemOpen
  \bibfield  {author} {\bibinfo {author} {\bibfnamefont {R.}~\bibnamefont
  {K\"{o}hler}}, \bibinfo {author} {\bibfnamefont {A.}~\bibnamefont
  {Tredicucci}}, \bibinfo {author} {\bibfnamefont {F.}~\bibnamefont {Beltram}},
  \bibinfo {author} {\bibfnamefont {H.~E.}\ \bibnamefont {Beere}}, \bibinfo
  {author} {\bibfnamefont {E.~H.}\ \bibnamefont {Linfield}}, \bibinfo {author}
  {\bibfnamefont {A.~G.}\ \bibnamefont {Davies}}, \bibinfo {author}
  {\bibfnamefont {D.~A.}\ \bibnamefont {Ritchie}}, \bibinfo {author}
  {\bibfnamefont {R.~C.}\ \bibnamefont {Iotti}}, \ and\ \bibinfo {author}
  {\bibfnamefont {F.}~\bibnamefont {Rossi}},\ }\href {<Go to
  ISI>://000175460200038} {\bibfield  {journal} {\bibinfo  {journal} {Nature}\
  }\textbf {\bibinfo {volume} {417}},\ \bibinfo {pages} {156} (\bibinfo {year}
  {2002})}\BibitemShut {NoStop}%
\bibitem [{\citenamefont {R\"{o}sch}\ \emph {et~al.}(2016)\citenamefont
  {R\"{o}sch}, \citenamefont {Scalari}, \citenamefont {Villares}, \citenamefont
  {Bosco}, \citenamefont {Beck},\ and\ \citenamefont {Faist}}]{DCMarkus}%
  \BibitemOpen
  \bibfield  {author} {\bibinfo {author} {\bibfnamefont {M.}~\bibnamefont
  {R\"{o}sch}}, \bibinfo {author} {\bibfnamefont {G.}~\bibnamefont {Scalari}},
  \bibinfo {author} {\bibfnamefont {G.}~\bibnamefont {Villares}}, \bibinfo
  {author} {\bibfnamefont {L.}~\bibnamefont {Bosco}}, \bibinfo {author}
  {\bibfnamefont {M.}~\bibnamefont {Beck}}, \ and\ \bibinfo {author}
  {\bibfnamefont {J.}~\bibnamefont {Faist}},\ }\href {<Go to
  ISI>://WOS:000375846600004} {\bibfield  {journal} {\bibinfo  {journal}
  {Applied Physics Letters}\ }\textbf {\bibinfo {volume} {108}},\ \bibinfo
  {pages} {171104} (\bibinfo {year} {2016})}\BibitemShut {NoStop}%
\bibitem [{\citenamefont {R\"{o}sch}\ \emph {et~al.}(2018)\citenamefont
  {R\"{o}sch}, \citenamefont {Beck}, \citenamefont {Suess}, \citenamefont
  {Bachmann}, \citenamefont {Unterrainer}, \citenamefont {Faist},\ and\
  \citenamefont {Scalari}}]{Octave}%
  \BibitemOpen
  \bibfield  {author} {\bibinfo {author} {\bibfnamefont {M.}~\bibnamefont
  {R\"{o}sch}}, \bibinfo {author} {\bibfnamefont {M.}~\bibnamefont {Beck}},
  \bibinfo {author} {\bibfnamefont {M.~J.}\ \bibnamefont {Suess}}, \bibinfo
  {author} {\bibfnamefont {D.}~\bibnamefont {Bachmann}}, \bibinfo {author}
  {\bibfnamefont {K.}~\bibnamefont {Unterrainer}}, \bibinfo {author}
  {\bibfnamefont {J.}~\bibnamefont {Faist}}, \ and\ \bibinfo {author}
  {\bibfnamefont {G.}~\bibnamefont {Scalari}},\ }\href {<Go to
  ISI>://WOS:000414650700013} {\bibfield  {journal} {\bibinfo  {journal}
  {Nanophotonics}\ }\textbf {\bibinfo {volume} {7}},\ \bibinfo {pages} {237}
  (\bibinfo {year} {2018})}\BibitemShut {NoStop}%
\bibitem [{\citenamefont {Villares}\ \emph {et~al.}(2014)\citenamefont
  {Villares}, \citenamefont {Hugi}, \citenamefont {Blaser},\ and\ \citenamefont
  {Faist}}]{DCFaist}%
  \BibitemOpen
  \bibfield  {author} {\bibinfo {author} {\bibfnamefont {G.}~\bibnamefont
  {Villares}}, \bibinfo {author} {\bibfnamefont {A.}~\bibnamefont {Hugi}},
  \bibinfo {author} {\bibfnamefont {S.}~\bibnamefont {Blaser}}, \ and\ \bibinfo
  {author} {\bibfnamefont {J.}~\bibnamefont {Faist}},\ }\href@noop {}
  {\bibfield  {journal} {\bibinfo  {journal} {Nature Communications}\ }\textbf
  {\bibinfo {volume} {5}},\ \bibinfo {pages} {5192} (\bibinfo {year}
  {2014})}\BibitemShut {NoStop}%
\bibitem [{\citenamefont {Yang}\ \emph {et~al.}(2016)\citenamefont {Yang},
  \citenamefont {Burghoff}, \citenamefont {Hayton}, \citenamefont {Gao},
  \citenamefont {Reno},\ and\ \citenamefont {Hu}}]{Yang16}%
  \BibitemOpen
  \bibfield  {author} {\bibinfo {author} {\bibfnamefont {Y.}~\bibnamefont
  {Yang}}, \bibinfo {author} {\bibfnamefont {D.}~\bibnamefont {Burghoff}},
  \bibinfo {author} {\bibfnamefont {D.~J.}\ \bibnamefont {Hayton}}, \bibinfo
  {author} {\bibfnamefont {J.~R.}\ \bibnamefont {Gao}}, \bibinfo {author}
  {\bibfnamefont {J.~L.}\ \bibnamefont {Reno}}, \ and\ \bibinfo {author}
  {\bibfnamefont {Q.}~\bibnamefont {Hu}},\ }\href@noop {} {\bibfield  {journal}
  {\bibinfo  {journal} {Optica}\ }\textbf {\bibinfo {volume} {3}},\ \bibinfo
  {pages} {499} (\bibinfo {year} {2016})}\BibitemShut {NoStop}%
\bibitem [{\citenamefont {Sterczewski}\ \emph {et~al.}(2019)\citenamefont
  {Sterczewski}, \citenamefont {Westberg}, \citenamefont {Yang}, \citenamefont
  {Burghoff}, \citenamefont {Reno}, \citenamefont {Hu},\ and\ \citenamefont
  {Wysocki}}]{WysockiDC}%
  \BibitemOpen
  \bibfield  {author} {\bibinfo {author} {\bibfnamefont {L.~A.}\ \bibnamefont
  {Sterczewski}}, \bibinfo {author} {\bibfnamefont {J.}~\bibnamefont
  {Westberg}}, \bibinfo {author} {\bibfnamefont {Y.}~\bibnamefont {Yang}},
  \bibinfo {author} {\bibfnamefont {D.}~\bibnamefont {Burghoff}}, \bibinfo
  {author} {\bibfnamefont {J.}~\bibnamefont {Reno}}, \bibinfo {author}
  {\bibfnamefont {Q.}~\bibnamefont {Hu}}, \ and\ \bibinfo {author}
  {\bibfnamefont {G.}~\bibnamefont {Wysocki}},\ }\href@noop {} {\bibfield
  {journal} {\bibinfo  {journal} {Optica}\ }\textbf {\bibinfo {volume} {6}},\
  \bibinfo {pages} {766} (\bibinfo {year} {2019})}\BibitemShut {NoStop}%
\bibitem [{\citenamefont {Li}\ \emph {et~al.}(2019{\natexlab{a}})\citenamefont
  {Li}, \citenamefont {Wan}, \citenamefont {Zhou}, \citenamefont {Liao},
  \citenamefont {Yang}, \citenamefont {Fu}, \citenamefont {Cao},\ and\
  \citenamefont {Li}}]{Lionchip}%
  \BibitemOpen
  \bibfield  {author} {\bibinfo {author} {\bibfnamefont {Z.}~\bibnamefont
  {Li}}, \bibinfo {author} {\bibfnamefont {W.}~\bibnamefont {Wan}}, \bibinfo
  {author} {\bibfnamefont {K.}~\bibnamefont {Zhou}}, \bibinfo {author}
  {\bibfnamefont {X.}~\bibnamefont {Liao}}, \bibinfo {author} {\bibfnamefont
  {S.}~\bibnamefont {Yang}}, \bibinfo {author} {\bibfnamefont {Z.}~\bibnamefont
  {Fu}}, \bibinfo {author} {\bibfnamefont {J.~C.}\ \bibnamefont {Cao}}, \ and\
  \bibinfo {author} {\bibfnamefont {H.}~\bibnamefont {Li}},\ }\href {\doibase
  10.1103/PhysRevApplied.12.044068} {\bibfield  {journal} {\bibinfo  {journal}
  {Physical Review Applied}\ }\textbf {\bibinfo {volume} {12}},\ \bibinfo
  {pages} {044068} (\bibinfo {year} {2019}{\natexlab{a}})}\BibitemShut
  {NoStop}%
\bibitem [{\citenamefont {Danylov}\ \emph {et~al.}(2015)\citenamefont
  {Danylov}, \citenamefont {Erickson}, \citenamefont {Light},\ and\
  \citenamefont {Waldman}}]{Schottky}%
  \BibitemOpen
  \bibfield  {author} {\bibinfo {author} {\bibfnamefont {A.}~\bibnamefont
  {Danylov}}, \bibinfo {author} {\bibfnamefont {N.}~\bibnamefont {Erickson}},
  \bibinfo {author} {\bibfnamefont {A.}~\bibnamefont {Light}}, \ and\ \bibinfo
  {author} {\bibfnamefont {J.}~\bibnamefont {Waldman}},\ }\href {<Go to
  ISI>://WOS:000364468600068} {\bibfield  {journal} {\bibinfo  {journal}
  {Optics Letters}\ }\textbf {\bibinfo {volume} {40}},\ \bibinfo {pages} {5090}
  (\bibinfo {year} {2015})}\BibitemShut {NoStop}%
\bibitem [{\citenamefont {Richter}\ \emph {et~al.}(2008)\citenamefont
  {Richter}, \citenamefont {Semenov}, \citenamefont {Pavlov}, \citenamefont
  {Mahler}, \citenamefont {Tredicucci}, \citenamefont {Beere}, \citenamefont
  {Ritchie}, \citenamefont {Il'in}, \citenamefont {Siegel},\ and\ \citenamefont
  {Hubers}}]{HEB}%
  \BibitemOpen
  \bibfield  {author} {\bibinfo {author} {\bibfnamefont {H.}~\bibnamefont
  {Richter}}, \bibinfo {author} {\bibfnamefont {A.~D.}\ \bibnamefont
  {Semenov}}, \bibinfo {author} {\bibfnamefont {S.~G.}\ \bibnamefont {Pavlov}},
  \bibinfo {author} {\bibfnamefont {L.}~\bibnamefont {Mahler}}, \bibinfo
  {author} {\bibfnamefont {A.}~\bibnamefont {Tredicucci}}, \bibinfo {author}
  {\bibfnamefont {H.~E.}\ \bibnamefont {Beere}}, \bibinfo {author}
  {\bibfnamefont {D.~A.}\ \bibnamefont {Ritchie}}, \bibinfo {author}
  {\bibfnamefont {K.~S.}\ \bibnamefont {Il'in}}, \bibinfo {author}
  {\bibfnamefont {M.}~\bibnamefont {Siegel}}, \ and\ \bibinfo {author}
  {\bibfnamefont {H.~W.}\ \bibnamefont {Hubers}},\ }\href {\doibase Artn 141108
  Doi 10.1063/1.2988896} {\bibfield  {journal} {\bibinfo  {journal} {Applied
  Physics Letters}\ }\textbf {\bibinfo {volume} {93}},\  (\bibinfo {year}
  {2008})}\BibitemShut {NoStop}%
\bibitem [{\citenamefont {Li}\ \emph {et~al.}(2017)\citenamefont {Li},
  \citenamefont {Wan}, \citenamefont {Tan}, \citenamefont {Fu}, \citenamefont
  {Wang}, \citenamefont {Zhou}, \citenamefont {Li}, \citenamefont {Wang},
  \citenamefont {Guo},\ and\ \citenamefont {Cao}}]{LiSR}%
  \BibitemOpen
  \bibfield  {author} {\bibinfo {author} {\bibfnamefont {H.}~\bibnamefont
  {Li}}, \bibinfo {author} {\bibfnamefont {W.-J.}\ \bibnamefont {Wan}},
  \bibinfo {author} {\bibfnamefont {Z.-Y.}\ \bibnamefont {Tan}}, \bibinfo
  {author} {\bibfnamefont {Z.-L.}\ \bibnamefont {Fu}}, \bibinfo {author}
  {\bibfnamefont {H.-X.}\ \bibnamefont {Wang}}, \bibinfo {author}
  {\bibfnamefont {T.}~\bibnamefont {Zhou}}, \bibinfo {author} {\bibfnamefont
  {Z.-P.}\ \bibnamefont {Li}}, \bibinfo {author} {\bibfnamefont
  {C.}~\bibnamefont {Wang}}, \bibinfo {author} {\bibfnamefont {X.-G.}\
  \bibnamefont {Guo}}, \ and\ \bibinfo {author} {\bibfnamefont {J.-C.}\
  \bibnamefont {Cao}},\ }\href {\doibase 10.1038/s41598-017-03787-6} {\bibfield
   {journal} {\bibinfo  {journal} {Scientific Reports}\ }\textbf {\bibinfo
  {volume} {7}},\ \bibinfo {pages} {3452} (\bibinfo {year} {2017})}\BibitemShut
  {NoStop}%
\bibitem [{\citenamefont {Hu}\ \emph {et~al.}(2018)\citenamefont {Hu},
  \citenamefont {Mizuguchi}, \citenamefont {Oe}, \citenamefont {Nitta},
  \citenamefont {Zhao}, \citenamefont {Minamikawa}, \citenamefont {Li},
  \citenamefont {Zheng},\ and\ \citenamefont {Yasui}}]{HuDCS2018}%
  \BibitemOpen
  \bibfield  {author} {\bibinfo {author} {\bibfnamefont {G.~Q.}\ \bibnamefont
  {Hu}}, \bibinfo {author} {\bibfnamefont {T.}~\bibnamefont {Mizuguchi}},
  \bibinfo {author} {\bibfnamefont {R.}~\bibnamefont {Oe}}, \bibinfo {author}
  {\bibfnamefont {K.}~\bibnamefont {Nitta}}, \bibinfo {author} {\bibfnamefont
  {X.}~\bibnamefont {Zhao}}, \bibinfo {author} {\bibfnamefont {T.}~\bibnamefont
  {Minamikawa}}, \bibinfo {author} {\bibfnamefont {T.}~\bibnamefont {Li}},
  \bibinfo {author} {\bibfnamefont {Z.}~\bibnamefont {Zheng}}, \ and\ \bibinfo
  {author} {\bibfnamefont {T.}~\bibnamefont {Yasui}},\ }\href {\doibase Artn
  11155 10.1038/S41598-018-29403-9} {\bibfield  {journal} {\bibinfo  {journal}
  {Scientific Reports}\ }\textbf {\bibinfo {volume} {8}},\ \bibinfo {pages}
  {11155} (\bibinfo {year} {2018})}\BibitemShut {NoStop}%
\bibitem [{\citenamefont {Jerez}\ \emph {et~al.}(2019)\citenamefont {Jerez},
  \citenamefont {Walla}, \citenamefont {Betancur}, \citenamefont
  {Martin-Mateos}, \citenamefont {de~Dios},\ and\ \citenamefont
  {Acedo}}]{JerezDCS2019}%
  \BibitemOpen
  \bibfield  {author} {\bibinfo {author} {\bibfnamefont {B.}~\bibnamefont
  {Jerez}}, \bibinfo {author} {\bibfnamefont {F.}~\bibnamefont {Walla}},
  \bibinfo {author} {\bibfnamefont {A.}~\bibnamefont {Betancur}}, \bibinfo
  {author} {\bibfnamefont {P.}~\bibnamefont {Martin-Mateos}}, \bibinfo {author}
  {\bibfnamefont {C.}~\bibnamefont {de~Dios}}, \ and\ \bibinfo {author}
  {\bibfnamefont {A.~P.}\ \bibnamefont {Acedo}},\ }\href {\doibase
  10.1364/Ol.44.000415} {\bibfield  {journal} {\bibinfo  {journal} {Optics
  Letters}\ }\textbf {\bibinfo {volume} {44}},\ \bibinfo {pages} {415}
  (\bibinfo {year} {2019})}\BibitemShut {NoStop}%
\bibitem [{\citenamefont {Wienold}\ \emph {et~al.}(2010)\citenamefont
  {Wienold}, \citenamefont {Schrottke}, \citenamefont {Giehler}, \citenamefont
  {Hey}, \citenamefont {Anders},\ and\ \citenamefont {Grahn}}]{Hybrid}%
  \BibitemOpen
  \bibfield  {author} {\bibinfo {author} {\bibfnamefont {M.}~\bibnamefont
  {Wienold}}, \bibinfo {author} {\bibfnamefont {L.}~\bibnamefont {Schrottke}},
  \bibinfo {author} {\bibfnamefont {M.}~\bibnamefont {Giehler}}, \bibinfo
  {author} {\bibfnamefont {R.}~\bibnamefont {Hey}}, \bibinfo {author}
  {\bibfnamefont {W.}~\bibnamefont {Anders}}, \ and\ \bibinfo {author}
  {\bibfnamefont {H.~T.}\ \bibnamefont {Grahn}},\ }\href {<Go to
  ISI>://WOS:000281153600013} {\bibfield  {journal} {\bibinfo  {journal}
  {Applied Physics Letters}\ }\textbf {\bibinfo {volume} {97}},\ \bibinfo
  {pages} {071113} (\bibinfo {year} {2010})}\BibitemShut {NoStop}%
\bibitem [{\citenamefont {Zhou}\ \emph {et~al.}(2019)\citenamefont {Zhou},
  \citenamefont {Li}, \citenamefont {Wan}, \citenamefont {Li}, \citenamefont
  {Liao},\ and\ \citenamefont {Cao}}]{ZhouAPL}%
  \BibitemOpen
  \bibfield  {author} {\bibinfo {author} {\bibfnamefont {K.}~\bibnamefont
  {Zhou}}, \bibinfo {author} {\bibfnamefont {H.}~\bibnamefont {Li}}, \bibinfo
  {author} {\bibfnamefont {W.~J.}\ \bibnamefont {Wan}}, \bibinfo {author}
  {\bibfnamefont {Z.~P.}\ \bibnamefont {Li}}, \bibinfo {author} {\bibfnamefont
  {X.~Y.}\ \bibnamefont {Liao}}, \ and\ \bibinfo {author} {\bibfnamefont
  {J.~C.}\ \bibnamefont {Cao}},\ }\href@noop {} {\bibfield  {journal} {\bibinfo
   {journal} {Applied Physics Letters}\ }\textbf {\bibinfo {volume} {114}},\
  \bibinfo {pages} {191106} (\bibinfo {year} {2019})}\BibitemShut {NoStop}%
\bibitem [{\citenamefont {Li}\ \emph {et~al.}(2015)\citenamefont {Li},
  \citenamefont {Laffaille}, \citenamefont {Gacemi}, \citenamefont {Apfel},
  \citenamefont {Sirtori}, \citenamefont {Leonardon}, \citenamefont
  {Santarelli}, \citenamefont {R\"{o}sch}, \citenamefont {Scalari},
  \citenamefont {Beck}, \citenamefont {Faist}, \citenamefont {H\"{a}nsel},
  \citenamefont {Holzwarth},\ and\ \citenamefont {Barbieri}}]{LiOE}%
  \BibitemOpen
  \bibfield  {author} {\bibinfo {author} {\bibfnamefont {H.}~\bibnamefont
  {Li}}, \bibinfo {author} {\bibfnamefont {P.}~\bibnamefont {Laffaille}},
  \bibinfo {author} {\bibfnamefont {D.}~\bibnamefont {Gacemi}}, \bibinfo
  {author} {\bibfnamefont {M.}~\bibnamefont {Apfel}}, \bibinfo {author}
  {\bibfnamefont {C.}~\bibnamefont {Sirtori}}, \bibinfo {author} {\bibfnamefont
  {J.}~\bibnamefont {Leonardon}}, \bibinfo {author} {\bibfnamefont
  {G.}~\bibnamefont {Santarelli}}, \bibinfo {author} {\bibfnamefont
  {M.}~\bibnamefont {R\"{o}sch}}, \bibinfo {author} {\bibfnamefont
  {G.}~\bibnamefont {Scalari}}, \bibinfo {author} {\bibfnamefont
  {M.}~\bibnamefont {Beck}}, \bibinfo {author} {\bibfnamefont {J.}~\bibnamefont
  {Faist}}, \bibinfo {author} {\bibfnamefont {W.}~\bibnamefont {H\"{a}nsel}},
  \bibinfo {author} {\bibfnamefont {R.}~\bibnamefont {Holzwarth}}, \ and\
  \bibinfo {author} {\bibfnamefont {S.}~\bibnamefont {Barbieri}},\ }\href
  {\doibase 10.1364/OE.23.033270} {\bibfield  {journal} {\bibinfo  {journal}
  {Optics Express}\ }\textbf {\bibinfo {volume} {23}},\ \bibinfo {pages}
  {33270} (\bibinfo {year} {2015})}\BibitemShut {NoStop}%
\bibitem [{\citenamefont {Gellie}\ \emph {et~al.}(2010)\citenamefont {Gellie},
  \citenamefont {Barbieri}, \citenamefont {Lampin}, \citenamefont {Filloux},
  \citenamefont {Manquest}, \citenamefont {Sirtori}, \citenamefont {Sagnes},
  \citenamefont {Khanna}, \citenamefont {Linfield}, \citenamefont {Davies},
  \citenamefont {Beere},\ and\ \citenamefont {Ritchie}}]{Gellie10}%
  \BibitemOpen
  \bibfield  {author} {\bibinfo {author} {\bibfnamefont {P.}~\bibnamefont
  {Gellie}}, \bibinfo {author} {\bibfnamefont {S.}~\bibnamefont {Barbieri}},
  \bibinfo {author} {\bibfnamefont {J.~F.}\ \bibnamefont {Lampin}}, \bibinfo
  {author} {\bibfnamefont {P.}~\bibnamefont {Filloux}}, \bibinfo {author}
  {\bibfnamefont {C.}~\bibnamefont {Manquest}}, \bibinfo {author}
  {\bibfnamefont {C.}~\bibnamefont {Sirtori}}, \bibinfo {author} {\bibfnamefont
  {I.}~\bibnamefont {Sagnes}}, \bibinfo {author} {\bibfnamefont {S.~P.}\
  \bibnamefont {Khanna}}, \bibinfo {author} {\bibfnamefont {E.~H.}\
  \bibnamefont {Linfield}}, \bibinfo {author} {\bibfnamefont {A.~G.}\
  \bibnamefont {Davies}}, \bibinfo {author} {\bibfnamefont {H.}~\bibnamefont
  {Beere}}, \ and\ \bibinfo {author} {\bibfnamefont {D.}~\bibnamefont
  {Ritchie}},\ }\href {<Go to ISI>://WOS:000283679200027} {\bibfield  {journal}
  {\bibinfo  {journal} {Optics Express}\ }\textbf {\bibinfo {volume} {18}},\
  \bibinfo {pages} {20799} (\bibinfo {year} {2010})}\BibitemShut {NoStop}%
\bibitem [{\citenamefont {Miao}\ \emph {et~al.}(2018)\citenamefont {Miao},
  \citenamefont {Zhu}, \citenamefont {Zhao}, \citenamefont {Zhan},\ and\
  \citenamefont {Yue}}]{RH1}%
  \BibitemOpen
  \bibfield  {author} {\bibinfo {author} {\bibfnamefont {X.~Y.}\ \bibnamefont
  {Miao}}, \bibinfo {author} {\bibfnamefont {J.}~\bibnamefont {Zhu}}, \bibinfo
  {author} {\bibfnamefont {K.}~\bibnamefont {Zhao}}, \bibinfo {author}
  {\bibfnamefont {H.~L.}\ \bibnamefont {Zhan}}, \ and\ \bibinfo {author}
  {\bibfnamefont {W.~Z.}\ \bibnamefont {Yue}},\ }\href {\doibase
  10.1177/0003702818772853} {\bibfield  {journal} {\bibinfo  {journal} {Applied
  Spectroscopy}\ }\textbf {\bibinfo {volume} {72}},\ \bibinfo {pages} {1040}
  (\bibinfo {year} {2018})}\BibitemShut {NoStop}%
\bibitem [{\citenamefont {Kim}\ \emph {et~al.}(2018)\citenamefont {Kim},
  \citenamefont {Cha}, \citenamefont {Roy}, \citenamefont {Kim},\ and\
  \citenamefont {Ahn}}]{RH2}%
  \BibitemOpen
  \bibfield  {author} {\bibinfo {author} {\bibfnamefont {H.~S.}\ \bibnamefont
  {Kim}}, \bibinfo {author} {\bibfnamefont {S.~H.}\ \bibnamefont {Cha}},
  \bibinfo {author} {\bibfnamefont {B.}~\bibnamefont {Roy}}, \bibinfo {author}
  {\bibfnamefont {S.}~\bibnamefont {Kim}}, \ and\ \bibinfo {author}
  {\bibfnamefont {Y.~H.}\ \bibnamefont {Ahn}},\ }\href@noop {} {\bibfield
  {journal} {\bibinfo  {journal} {Optics Express}\ }\textbf {\bibinfo {volume}
  {26}},\ \bibinfo {pages} {33575} (\bibinfo {year} {2018})}\BibitemShut
  {NoStop}%
\bibitem [{\citenamefont {Li}\ \emph {et~al.}(2014)\citenamefont {Li},
  \citenamefont {Manceau}, \citenamefont {Andronico}, \citenamefont {Jagtap},
  \citenamefont {Sirtori}, \citenamefont {Li}, \citenamefont {Linfield},
  \citenamefont {Davies},\ and\ \citenamefont {Barbieri}}]{LiCC}%
  \BibitemOpen
  \bibfield  {author} {\bibinfo {author} {\bibfnamefont {H.}~\bibnamefont
  {Li}}, \bibinfo {author} {\bibfnamefont {J.~M.}\ \bibnamefont {Manceau}},
  \bibinfo {author} {\bibfnamefont {A.}~\bibnamefont {Andronico}}, \bibinfo
  {author} {\bibfnamefont {V.}~\bibnamefont {Jagtap}}, \bibinfo {author}
  {\bibfnamefont {C.}~\bibnamefont {Sirtori}}, \bibinfo {author} {\bibfnamefont
  {L.~H.}\ \bibnamefont {Li}}, \bibinfo {author} {\bibfnamefont {E.~H.}\
  \bibnamefont {Linfield}}, \bibinfo {author} {\bibfnamefont {A.~G.}\
  \bibnamefont {Davies}}, \ and\ \bibinfo {author} {\bibfnamefont
  {S.}~\bibnamefont {Barbieri}},\ }\href {\doibase Artn 241102
  10.1063/1.4884056} {\bibfield  {journal} {\bibinfo  {journal} {Applied
  Physics Letters}\ }\textbf {\bibinfo {volume} {104}},\ \bibinfo {pages}
  {241102} (\bibinfo {year} {2014})}\BibitemShut {NoStop}%
\bibitem [{\citenamefont {Li}\ \emph {et~al.}(2019{\natexlab{b}})\citenamefont
  {Li}, \citenamefont {Yan}, \citenamefont {Wan}, \citenamefont {Zhou},
  \citenamefont {Zhou}, \citenamefont {Li}, \citenamefont {Cao}, \citenamefont
  {Yu}, \citenamefont {Zhang}, \citenamefont {Li}, \citenamefont {Nan},
  \citenamefont {He},\ and\ \citenamefont {Zeng}}]{LiAS}%
  \BibitemOpen
  \bibfield  {author} {\bibinfo {author} {\bibfnamefont {H.}~\bibnamefont
  {Li}}, \bibinfo {author} {\bibfnamefont {M.}~\bibnamefont {Yan}}, \bibinfo
  {author} {\bibfnamefont {W.}~\bibnamefont {Wan}}, \bibinfo {author}
  {\bibfnamefont {T.}~\bibnamefont {Zhou}}, \bibinfo {author} {\bibfnamefont
  {K.}~\bibnamefont {Zhou}}, \bibinfo {author} {\bibfnamefont {Z.}~\bibnamefont
  {Li}}, \bibinfo {author} {\bibfnamefont {J.}~\bibnamefont {Cao}}, \bibinfo
  {author} {\bibfnamefont {Q.}~\bibnamefont {Yu}}, \bibinfo {author}
  {\bibfnamefont {K.}~\bibnamefont {Zhang}}, \bibinfo {author} {\bibfnamefont
  {M.}~\bibnamefont {Li}}, \bibinfo {author} {\bibfnamefont {J.}~\bibnamefont
  {Nan}}, \bibinfo {author} {\bibfnamefont {B.}~\bibnamefont {He}}, \ and\
  \bibinfo {author} {\bibfnamefont {H.}~\bibnamefont {Zeng}},\ }\href {\doibase
  10.1002/advs.201900460} {\bibfield  {journal} {\bibinfo  {journal} {Advanced
  Science}\ }\textbf {\bibinfo {volume} {6}},\ \bibinfo {pages} {1900460}
  (\bibinfo {year} {2019}{\natexlab{b}})}\BibitemShut {NoStop}%
\bibitem [{\citenamefont {Barbieri}\ \emph {et~al.}(2011)\citenamefont
  {Barbieri}, \citenamefont {Ravaro}, \citenamefont {Gellie}, \citenamefont
  {Santarelli}, \citenamefont {Manquest}, \citenamefont {Sirtori},
  \citenamefont {Khanna}, \citenamefont {Linfield},\ and\ \citenamefont
  {Davies}}]{BarbieriNP}%
  \BibitemOpen
  \bibfield  {author} {\bibinfo {author} {\bibfnamefont {S.}~\bibnamefont
  {Barbieri}}, \bibinfo {author} {\bibfnamefont {M.}~\bibnamefont {Ravaro}},
  \bibinfo {author} {\bibfnamefont {P.}~\bibnamefont {Gellie}}, \bibinfo
  {author} {\bibfnamefont {G.}~\bibnamefont {Santarelli}}, \bibinfo {author}
  {\bibfnamefont {C.}~\bibnamefont {Manquest}}, \bibinfo {author}
  {\bibfnamefont {C.}~\bibnamefont {Sirtori}}, \bibinfo {author} {\bibfnamefont
  {S.~P.}\ \bibnamefont {Khanna}}, \bibinfo {author} {\bibfnamefont {E.~H.}\
  \bibnamefont {Linfield}}, \ and\ \bibinfo {author} {\bibfnamefont {A.~G.}\
  \bibnamefont {Davies}},\ }\href {<Go to ISI>://WOS:000290014900022}
  {\bibfield  {journal} {\bibinfo  {journal} {Nature Photonics}\ }\textbf
  {\bibinfo {volume} {5}},\ \bibinfo {pages} {306} (\bibinfo {year}
  {2011})}\BibitemShut {NoStop}%
\end{thebibliography}%

\end{document}